\documentclass[conference,linesnumbered,ruled,vlined]{IEEEtran}
\usepackage[latin9]{inputenc}
\usepackage{wrapfig}
\usepackage{algorithm2e}
\usepackage{amsmath}
\usepackage{amssymb}
\usepackage{graphicx}

\makeatletter
\IEEEoverridecommandlockouts

\usepackage{cite}
\usepackage{amsfonts}
\usepackage{textcomp}
\usepackage{xcolor}
\def\BibTeX{{\rm B\kern-.05em{\sc i\kern-.025em b}\kern-.08em
    T\kern-.1667em\lower.7ex\hbox{E}\kern-.125emX}}

\usepackage{algpseudocode}

\algnewcommand\algorithmicforeach{\textbf{for each}}
\algdef{S}[FOR]{ForEach}[1]{\algorithmicforeach\ #1\ \algorithmicdo}

\makeatother

\begin{document}
\title{On the use of associative memory in Hopfield networks designed to
solve propositional satisfiability problems}
\author{\IEEEauthorblockN{Natalya Weber} \IEEEauthorblockA{\textit{Embodied Cognitive Science Unit} \\
 \textit{\emph{Okinawa Institute of Science and }}\\
 \textit{\emph{Technology Graduate University}}\\
 Okinawa, Japan\\
 ORCID: 0000-0002-1955-3612} \and \IEEEauthorblockN{Werner Koch}\IEEEauthorblockA{\textit{Independent Scholar}\\
 Dresden, Germany \\
 ORCID: \\
0000-0001-7246-0434}\and\IEEEauthorblockN{Ozan Erdem} \IEEEauthorblockA{\textit{Independent Scholar}\\
 Toronto, Canada \\
 ORCID: \\
0009-0003-7844-4832}\and\IEEEauthorblockN{Tom Froese} \IEEEauthorblockA{\textit{Embodied Cognitive Science Unit} \\
 \textit{\emph{Okinawa Institute of Science and }}\\
 \textit{\emph{Technology Graduate University}}\\
 Okinawa, Japan\\
 tom.froese@oist.jp}}
\maketitle
\begin{abstract}
Hopfield networks are an attractive choice for solving many types
of computational problems because they provide a biologically plausible
mechanism. The Self-Optimization (SO) model adds to the Hopfield network
by using a biologically founded Hebbian learning rule, in combination
with repeated network resets to arbitrary initial states, for optimizing
its own behavior towards some desirable goal state encoded in the
network. In order to better understand that process, we demonstrate
first that the SO model can solve concrete combinatorial problems
in SAT form, using two examples of the Liars problem and the map coloring
problem. In addition, we show how under some conditions critical information
might get lost forever with the learned network producing seemingly
optimal solutions that are in fact inappropriate for the problem it
was tasked to solve. What appears to be an undesirable side-effect
of the SO model, can provide insight into its process for solving
intractable problems. 
\end{abstract}


\IEEEoverridecommandlockouts
\IEEEpubid{\begin{minipage}
{\textwidth}\ \\[12pt]\copyright\ 2023 IEEE. Personal use of this material is permitted. Permission from IEEE must be obtained for all other uses, in any current or future media, including reprinting/republishing this material for advertising or promotional purposes, creating new collective works, for resale or redistribution to servers or lists, or reuse of any copyrighted component of this work in other works.
\end{minipage}}

\section{Introduction\label{sec:Introduction}}

The combination of domain knowledge and centralized control is an
effective solution to a broad class of optimization problems. However,
in the case of complex adaptive systems, the system's control tends
to be distributed and it is often unclear what the most appropriate
trajectory is and even the form of the optimal solution may simply
be unknown. This is the case for many kinds of biological systems,
but also social systems, that tend to be capable of giving rise to
creative solutions even under novel circumstances. Such a complex
adaptive system cannot necessarily rely on the availability of error
or reward signals to improve its behavior, which raises the intriguing
question of what other, more minimal mechanisms could be available.

\IEEEpubidadjcol

A particularly interesting model of a distributed complex adaptive
system is the Self-Optimization (SO) model. It is a simple model comprised
of a Hopfield network (HN) \cite{hopfield_neural_1982} and Hebbian
learning \cite{hebb_organization_1949}, two biologically founded
\cite{pulvermuller_biological_2021} and well established mechanisms.
Already in 1985 Hopfield and Tank showed that if an optimization problem
is formulated in terms of desired optima subject to constraints (i.e.
the connections of the network correspond to the constraints of the
problem), the natural dynamics of the system is to converge to a stable
state that will correspond to a locally optimal solution to that problem
with the least violated constraints \cite{hopfield_neural_1985}.
The fascinating thing about the SO model is that under certain conditions,
the combination of Hebbian learning with the dynamics of the HN allows
the system to form an associative memory of its own behaviour and
change its own dynamics (hence ``self'') to enhance its ability
to find configurations that minimize the constraints between system's
variables, and ``find solutions that are better than any solutions
found before the application of such learning (i.e. \emph{true optimisation})''
\cite[emphasis added]{watson_collective_2023}.

\IEEEpubidadjcol

Closely related to the field of engineering is the scientific discipline
of propositional satisfiability (SAT) problems. The goal in a SAT
problem is to determine whether a given logical formula can be made
true by assigning appropriate truth values to its variables. Many
important real-world problems in different scientific fields can be
naturally expressed as MaxSAT \cite{biere_handbook_2009}: routing
and scheduling problems in industrial engineering, software and hardware
debugging in computer science and computer engineering, different
problems of bioinformatics in biological sciences, just to name a
few. It was previously mentioned \cite{watson_optimization_2011}
that the initial weights of the HN network in an optimization framework
represent a weighted-Max-2-SAT problem, but it was never actually
shown how one would start from a SAT problem in question and use the
SO model to solve it (an analogous model to that of SO was used before
to solve a concrete problem \cite{power_distributed_2019}, but not
in the form of a SAT problem on which we expand subsequently). This
poses an obstacle for researches coming from different fields to understand
how they would be able to apply the model into their research directly.

Thus our goal in this work is twofold. First we want to present a
method for converting any SAT problem to the weights of the Hopfield
network such that researchers from any related field could try their
own problems on this model. Second, we want to show how by using a
concrete problem we can further expand our knowledge about the SO
model in general, by being able to answer questions that we could
not answer in the abstract case.

In this work we built upon and combine work from different areas.
Accordingly, we want to point out the similarities and differences
in each case: 
\begin{itemize}
\item Two different research lines, \cite{lima_satyrus_2005} and \cite{mohd_shareduwan_mohd_kasihmuddin_discrete_2018},
performed on opposite sides of the globe, use Hopfield networks to
solve SAT problems. The former uses the method described in \cite{pinkas_symmetric_1991}
for translating the list of clauses of a SAT problem into the energy
function, and the latter uses the method described in \cite{abdullah_logic_1992}
to further translate the energy function into the weights for the
Hopfield network. Neither \cite{lima_satyrus_2005} or \cite{mohd_shareduwan_mohd_kasihmuddin_discrete_2018}
use learning in their simulations. In the current work, we use the
\cite{abdullah_logic_1992} method and our model uses Hebbian learning. 
\item It should be pointed out that Hebbian learning in the context of logic
of neural networks was investigated before in \cite{ahmad_tajuddin_wan_abdullah_logic_1993},
but the goal of that investigation was not to improve the problem
solving of the system, but to perform a reverse analysis - given the
obtained weights through learning, the question was to obtain the
logical clauses acquired by the system. In comparison, in the current
work Hebbian learning is used along with periodic resets of the system
in order to enhance the ability of the system to find configurations
that minimize the violations of constraints. 
\item As mentioned above, analogous dynamics to that of the SO model is
present in the coevolution in ecological networks \cite{power_distributed_2019},
and in that work the author also shows how the constraint satisfaction
behavior of their model enables it to solve Sudoku puzzles. Different
from that work, in the current work we want to re-introduce to the
community an already relatively old method, that didn't get enough
exposure, for solving any SAT problem on the SO model so that researchers
from different disciplines would be able to apply the SO model for
their needs. In addition, we discuss the implications of breaking
constraints on the end result of the solution, something that was
never discussed before. 
\end{itemize}
The rest of the paper is structured as follows: in Sec.~\ref{sec:Background}
we give a short background to the research preceding the development
of the ``Abdullah method'' used in this work and further described
in Sec.~\ref{sec:Conversion}. Section~\ref{sec:Conversion} describes
how a SAT problem is translated to the HN weights, which can then
be used in the SO simulation. We provide examples for two classic
problems - the Liars problem and map coloring, which is a special
case of graph coloring. Section~\ref{sec:Results} is broken into
two parts. We first show in Sec.~\ref{subsec:Solvable-problems}
that the SO model successfully solves both of the problems when a
satisfiable solution exists. Then in Sec.~\ref{subsec:South America}
we discuss what it means to break constraints in the context of an
unsatisfiable problem. The problem examined is that of coloring a
geographical map with just 2 colors where a minimum of 4 colors is
required. Finally, in Section~\ref{sec:Discussion} we draw the conclusions
of this work and discuss the various paths for future research.

\section{Background: high-order Hopfield networks and logic for problem solving\label{sec:Background}}

Following Hopfield's groundbreaking papers \cite{hopfield_neural_1982,hopfield_neural_1985},
several research lines \cite{sejnowski_higherorder_1986,baldi_number_1987,abbott_storage_1987}
generalized the energy function of the HN to include high-order terms.
Here we adopt the following notation: 
\begin{align}
E^{(k)}\left(t\right)= & -\frac{1}{k}\sum_{\gamma_{1}}^{N}\sum_{\gamma_{2}}^{N}\cdots\sum_{\gamma_{k}}^{N}W_{\gamma_{1}\cdots\gamma_{k}}^{(k)}\prod_{i=1}^{k}s_{\gamma_{i}}\left(t\right)\nonumber \\
 & -\frac{1}{k-1}\sum_{\gamma_{1}}^{N}\cdots\sum_{\gamma_{k-1}}^{N}W_{\gamma_{1}\cdots\gamma_{k-1}}^{(k-1)}\prod_{i=1}^{k-1}s_{\gamma_{i}}\left(t\right)-\cdots\nonumber \\
 & -\frac{1}{2}\sum_{i}^{N}\sum_{j}^{N}W_{ij}^{(2)}s_{i}\left(t\right)s_{j}\left(t\right)-\sum_{i}^{N}W_{i}^{(1)}s_{i}\left(t\right)-c,\label{eq:E_high-order}
\end{align}
where $k$ is the order of the energy function, $W_{\gamma_{1}\cdots\gamma_{k}}^{(k)}$
is a $k$-th order tensor of size $N^{k}$, symmetric on all pairs
of indices $\gamma_{i}$. The $s_{i}$ are bipolar discrete elements
of the system's state vector $\mathbf{S}=\left\{ s_{1}\left(t\right),...,s_{N}\left(t\right)\right\} $
of size $N$, and $c$ is a constant. The standard HN energy function
is then a special case of \eqref{eq:E_high-order} with $k=2$: 
\begin{equation}
E\left(t\right)=-\frac{1}{2}\sum_{i}^{N}\sum_{j}^{N}W_{ij}^{\left(2\right)}s_{i}\left(t\right)s_{j}\left(t\right)-\sum_{i}^{N}W_{i}^{(1)}s_{i}\left(t\right)-c.\label{eq:E_HN}
\end{equation}

In parallel to researchers in the fields of physics and neuroscience
expanding their knowledge on neural networks, researchers in the field
of artificial intelligence (AI) started working on developing logic
programming for problem solving \cite{kowalski_logic_1979-1}. Then
in 1991, \cite{pinkas_symmetric_1991} expanded this work by showing
an equivalence between the search problem of propositional logic satisfiability
and the problem of minimizing the energy function \eqref{eq:E_high-order}.
Building on that work, \cite{abdullah_logic_1992} presented a method
to compute the synaptic weights of the network, which correspond to
the propositional logic embedded in the system. The latter, termed
later \cite{sathasivam_logic_2008} as the ``Abdullah method'',
is the method adopted in the current work and is presented in Sec.~\ref{sec:Conversion}.

\section{SAT Problem conversion to Hopfield network weights\label{sec:Conversion}}

The satisfiability problem in propositional logic (SAT) is a combinatorial
problem of deciding for a given propositional formula $\Phi$, whether
there exists an assignment of truth values to the propositional variables
appearing in $\Phi$ under which the formula $\Phi$ evaluates to
``true''. Satisfying assignments are called ``models'' of $\Phi$
and form the solutions of the respective instance of SAT.

In the following we provide two different classical examples of SAT
problems - the Liars problem and the map coloring problem.

\subsection{Liars problem\label{subsec:Liars-problem}}

Imagine a room with $N=4$ people: Alice, Bob, Cal, and Dan. Some
(or all) of them can make statements, for example: 
\begin{quote}
Alice says ``Dan is a liar.'', Dan says ``Bob is a truth-teller'',
Cal says ``Bob is a liar''. 
\end{quote}
The assumption is that each person can either be a liar (always lying)
or a truth-teller (always telling the truth). The problem is to find
whether a valid assignment of ``Liar'' or ``Truth-teller'' for
each person exists given the statements above. These facts are represented
by a vector of states $\mathbf{S}=\left\{ s_{1},\ldots,s_{N}\right\} $
with the fact that the $i$-th person is a liar or a truth teller
represented by the state $s_{i}\in\left\{ 0,1\right\} ,\,i\in\left[1,N\right]$,
where $s_{i}=1$ denotes the person $i$ being a Truth-teller, and
$s_{i}=0$ denotes the person $i$ being a Liar. Using this notation
we can translate the above statements into a \emph{knowledge base,
}which represents the constraints for the liars and truth-tellers
problem\footnote{For readers without a background in SAT we recommend the following
two \cite{neller_clue_2008,erdem_encoding_2019} online resources
for short, but to the point very clear description of SAT and its
uses. For a detailed mathematical description see \cite{biere_handbook_2009}.}: 
\begin{equation}
\left\{ s_{1}\Leftrightarrow\lnot s_{4},s_{4}\Leftrightarrow s_{2},s_{3}\Leftrightarrow\lnot s_{2}\right\} .\label{eq:Liars_N4}
\end{equation}

Most SAT algorithms operate on propositional formulae in Conjunctive
Normal Form (CNF), which is as a \emph{conjunction of disjunctions
of literals. }Converting \eqref{eq:Liars_N4} to a CNF gives: 
\begin{align}
\Phi= & \left(\lnot s_{1}\lor\lnot s_{4}\right)\land\left(s_{4}\lor s_{1}\right)\land\left(\lnot s_{4}\lor s_{2}\right)\nonumber \\
 & \land\left(\lnot s_{2}\lor s_{4}\right)\land\left(\lnot s_{3}\lor\lnot s_{2}\right)\land\left(s_{2}\lor s_{3}\right).\label{eq:Liars_N4_cnf}
\end{align}

Next we consider another example and then in Sec.~\ref{subsec:Abdullah-method}
we show how to convert \eqref{eq:Liars_N4_cnf} to the weights of
a Hopfield network.

\subsection{Map coloring\label{subsec:Map-coloring} }

The map coloring problem is a special case of a graph coloring problem.
It is known that any map can be colored with just four colors \cite{mackenzie_slaying_1999},
however some maps might require just three or two colors. In this
case the state of the map is represented by a matrix $\mathbf{S}=\left\{ s_{1}^{1},s_{1}^{2},\ldots,s_{n}^{M}\right\} ,\,i\in\left[1,n\right],\,j\in\left[1,M\right]$
with $s_{i}^{j}=1$ denoting the region $i$ being colored with the
color $j$. Given the borders between all regions on the map, the
goal is to color all regions by distinct colors such that no two bordering
regions have the same color. Given $N=nM$ states for $n$ regions,
$M$ colors, and the border adjacency matrix $B$ with the elements
$b_{ii'}=1$ denoting region $i$ and region $i'$ sharing a border
and $b_{ii'}=0$ otherwise, the map coloring problem can be summarized
into the three sets of constraints: 
\begin{enumerate}
\item Each region has to be colored: 
\begin{equation}
\varphi_{1}=\bigwedge_{i=1}^{n}\left(\bigvee_{j=1}^{M}\left(s_{i}^{j}\right)\right)\,.\label{eq:Clauses - All-colored}
\end{equation}
\item A region cannot be two distinct colors at the same time: 
\begin{equation}
\varphi_{2}=\bigwedge_{i=1}^{n}\bigwedge_{j=1}^{M}\bigwedge_{j'\neq j}^{M}\left(\lnot\left(s_{i}^{j}\land s_{i}^{j'}\right)\right)\,.\label{eq:Clauses-all-proper-colors}
\end{equation}
\item Regions that share a border should have a different color: 
\end{enumerate}
\begin{equation}
\varphi_{3}=\bigwedge_{i=1}^{n}\bigwedge_{i'\neq i}^{n}\bigwedge_{j=1}^{M}\left(\lnot\left(s_{i}^{j}\land s_{i'}^{j}\right)\lor\lnot b_{ii'}\right)\,.\label{eq:Clauses-borders}
\end{equation}

Taking the sets of clauses \eqref{eq:Clauses - All-colored}, \eqref{eq:Clauses-all-proper-colors},
and \eqref{eq:Clauses-borders} together gives the formula for the
map coloring problem: 
\begin{align}
\Phi= & \varphi_{1}\land\varphi_{2}\land\varphi_{3}.\label{eq:MapColor_cnf}
\end{align}

Already for $n=4$ regions, and $M=2$ colors the propositional formula
$\Phi$ in \eqref{eq:MapColor_cnf} will have 20 clauses so it is
too lengthy to present in this paper, but we provide the full analytical
derivation in \cite{weber_so_2023}. In the next section we show how
to convert \eqref{eq:Liars_N4_cnf} and \eqref{eq:MapColor_cnf} to
the weights of a Hopfield network. 

\subsection{Using the Abdullah method to translate a logical formula to weights
of the Hopfield network\label{subsec:Abdullah-method}}

According to \cite{abdullah_logic_1992,ahmad_tajuddin_wan_abdullah_logic_1993}
determining the states $\mathbf{S}$ that will satisfy $\Phi$ is
equivalent to a combinatorial minimization of the cost function $E_{\lnot\Phi}$
of the \emph{inconsistency} $\lnot\Phi$. The value of $E_{\lnot\Phi}$
depends on the number of clauses satisfied by the model, such that
the more clauses are unsatisfied, the bigger the value of $E_{\lnot\Phi}$,
and $E_{\lnot\Phi}=0$ denoting that $\lnot\Phi$ evaluates to ``False'',
which means that $\Phi$ evaluates to ``True'' and a state $\mathbf{S}$
that satisfies all the constraints was found. The cost function $E_{\lnot\Phi}$
is given by a sum over one term each per negated clause in the logical
formula $\Phi$. For example, given formula \eqref{eq:Liars_N4_cnf},
the inconsistency $\lnot\Phi$ is 
\begin{align}
\lnot\Phi= & \left(s_{1}\land s_{4}\right)\lor\left(\lnot s_{4}\land\lnot s_{1}\right)\lor\left(s_{4}\land\lnot s_{2}\right)\nonumber \\
 & \land\left(s_{2}\land\lnot s_{4}\right)\lor\left(s_{3}\land s_{2}\right)\lor\left(\lnot s_{2}\land\lnot s_{3}\right).\label{eq:Liars_dnf}
\end{align}

The literals $s_{i}$ and $\lnot s_{i}$ in \eqref{eq:Liars_dnf}
are mapped to the terms $\frac{1}{2}\left(1+s_{i}\right)$ and $\frac{1}{2}\left(1-s_{i}\right)$,
respectively. The entire disjunction of conjuctions is subsequently
turned into a sum of products of such terms. In the example given
above, the corresponding cost function $E_{\lnot\Phi}$ takes the
form 
\begin{align}
E_{\lnot\Phi}= & \frac{1}{4}\left(1+s_{1}\right)\left(1+s_{4}\right)+\frac{1}{4}\left(1-s_{4}\right)\left(1-s_{1}\right)\nonumber \\
+ & \frac{1}{4}\left(1+s_{4}\right)\left(1-s_{2}\right)+\frac{1}{4}\left(1+s_{2}\right)\left(1-s_{4}\right)\nonumber \\
+ & \frac{1}{4}\left(1+s_{3}\right)\left(1+s_{2}\right)+\frac{1}{4}\left(1-s_{2}\right)\left(1-s_{3}\right)\nonumber \\
= & \frac{1}{2}\left(-s_{2}s_{3}-s_{1}s_{4}+s_{2}s_{4}\right)+\frac{3}{2}.\label{eq:Liars_cost}
\end{align}

We obtain the values of connections $W_{ij}^{\left(2\right)}$, the
biases $W_{i}^{(1)}$, and $c$ by comparing the cost function $E_{\lnot\Phi}$
\eqref{eq:Liars_cost} term by term with the energy function \eqref{eq:E_HN}.
Similarly, if the $E_{\lnot\Phi}$ had higher order products, we would
need to compare it to a higher order energy function \eqref{eq:E_high-order}.

Figure~\ref{fig:Liars-Checker-all}a and Figure~\ref{fig:Liars-Checker-all}d
show the connections $W_{ij}^{\left(2\right)}$ for the Liars problem
with $N=50$ people and for a map coloring problem in a form of a
checkerboard that has $n=8\times8=64$ tiles that need to be colored
by $M=2$ distinct colors, thus having $N=64\times2=128$ states.

Once we have the initial weight matrix $\mathbf{W}_{0}^{(2)}$, comprised
of the elements $W_{ij}^{\left(2\right)}$, we can use the SO model
to find the states $\mathbf{S}$ that will satisfy $\Phi$. Since
the mechanism \cite{watson_transformations_2011-1} and the implementation
\cite{weber_scaling_2022} of the SO model were previously covered
in length we do not expand on them here. The two examples serve to
show how different the weight connections of the two problems are,
which will affect the SO procedure as will be shown in the next section.

In the next section we present the results of running the SO model
for three different scenarios: in Sec.~\ref{subsec:Solvable-problems}
we show results for the the Liars problem, and the map problem of
a checkerboard, and in Sec.~\ref{subsec:South America} we show the
results for coloring the map of South America.

\section{Results\label{sec:Results}}

For all the results in this section we use the ``on-the-fly'' implementation
developed in previous work \cite{weber_scaling_2022}, and the algorithm
was implemented as a compiled FORTRAN module to be loaded from Python.
The shapes comprising the map of South America in Sec.~\ref{subsec:South America}
were obtained from the CShapes 2.0 Dataset \cite{schvitz_mapping_2022}.
The adjacency for the map coloring problem was determined geometrically
from the country map shapes, and the statements for the Liars problem
of the type described in Sec.~\ref{subsec:Liars-problem} were chosen
randomly. The clauses for each problem were generated by a dedicated
Python code available in \cite{weber_so_2023}. From the thus computed
clauses, the initial weights $\mathbf{W}_{0}$ for each problem were
obtained using the procedure described in Sec.~\ref{subsec:Abdullah-method}.

We first present in Sec.~\ref{subsec:Solvable-problems} that the
SO model successfully solves the two different instances of the Liars
problem and coloring of a checkerboard. Then in Sec.~\ref{subsec:South America}
we use the problem of coloring the map of South America and to show
what happens when certain constraints are broken and what are their
implications for the problem.

\subsection{Solvable problems\label{subsec:Solvable-problems}}

\begin{figure}
\vspace*{-3mm}
 \includegraphics{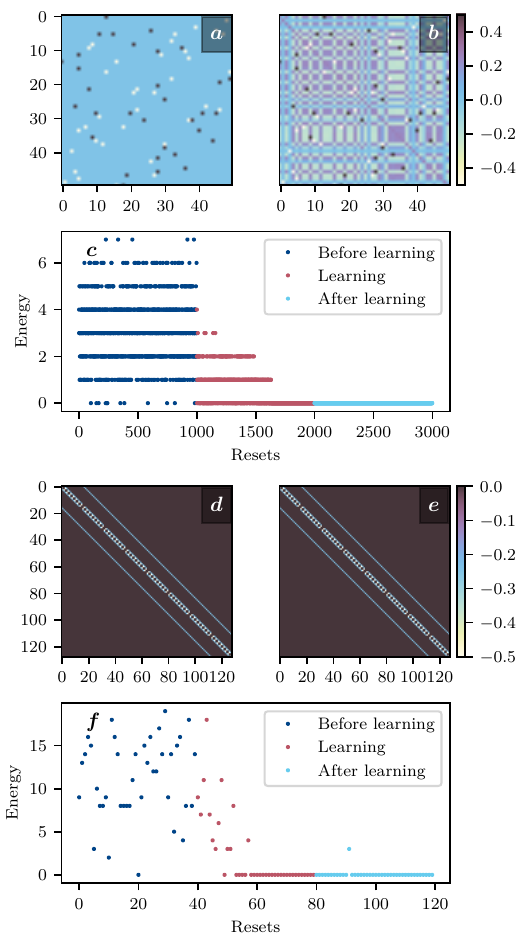} \caption{Simulation results for (a-c) the Liars problem ($N=50$ people, $34$
statements, learning rate $\alpha=2.5\times10^{-7}$, $20N$ steps),
and (d-f) the Checkerboard map coloring problem ($N=2\cdot64=128$
states, $\alpha=8\times10^{-21}$, $10N$ steps). (a),(d) The initial
weights $\mathbf{W}_{0}$ derived from the cost function $E_{\protect\lnot\Phi}$
comparison with \eqref{eq:E_HN}, (b),(e) the weights $\mathbf{W}$
after learning, (c),(f) the energy at the end of convergence for a
set without learning (resets 1--1000 and 1--40, blue), during learning
(1001--2000 and 41--80, red), and after learning (2001-3000 and
81-120, light blue) for the two problems, respectively.\label{fig:Liars-Checker-all}}
\vspace*{-3mm}
 
\end{figure}

The results for the two different instances of the Liars problem and
the map coloring problem of a checkerboard are shown in Fig.~\ref{fig:Liars-Checker-all}a-c
and Fig.~\ref{fig:Liars-Checker-all}d-f, respectively. We can see
that for both of the problems the SO model converges to an energy
$E=0$, denoting that a state $\mathbf{S}$ that satisfies all the
constraints was found for both cases. 

\begin{wrapfigure}[8]{r}{0.13\textwidth}%
 \centering \includegraphics{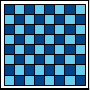} \caption{Colored checkerboard from the learned state $\mathbf{S}$.\label{fig:Checkerboard}}
\end{wrapfigure}

This is further exemplified by the correct coloring of the checkerboard
in Fig.~\ref{fig:Checkerboard}. We can see from the two figures
that the system in the checkerboard problem requires far fewer resets
and a learning rate more than 10 orders lower for convergence, despite
the fact that it has more nodes than in the Liars problem ($N=128$
tiles compared to $N=50$ people). This is because in the case of
the checkerboard, there is a band-diagonal structure in the constraints
of the problem. It is a very easy problem and just requires the barest
of nudges to settle into the correct configuration. In fact, with
larger learning rates, the system converges early to an incorrect
configuration without being able to escape it later on. Structure
in the constraints is known to improve the chances of the SO model
of finding a solution \cite{watson_transformations_2011-1}. Such
structure is missing for the Liars problem, making it difficult for
the system to find the proper solution. We will come back to this
point in Sec.~\ref{sec:Discussion}.

\subsection{Coloring the map of South America with 2 colors\label{subsec:South America}}

In this section we show that one can use the SO model%

for more difficult maps than just a checkerboard, specifically the
map of South America. As mentioned, coloring the map of South America
requires four colors, and here we use just two, so in this context
it is an unsolvable problem. 
\begin{figure*}
\includegraphics{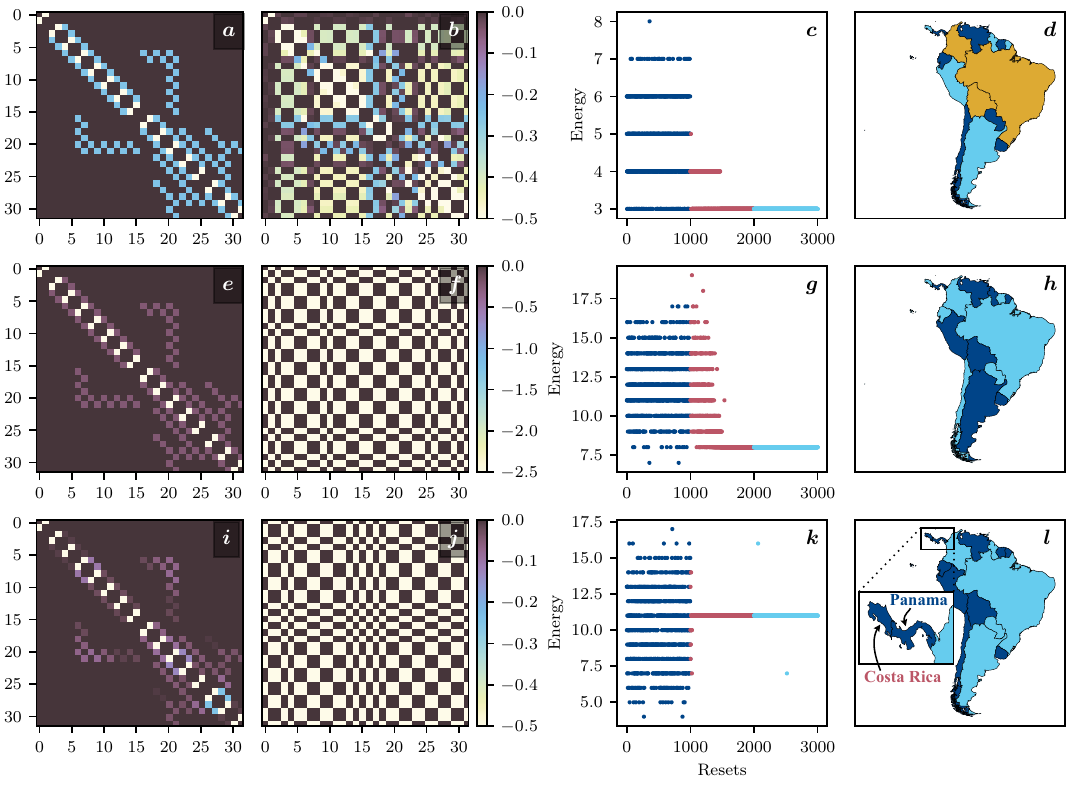}

\caption{Self-optimization simulation results for coloring the map of South
America ($N=2\cdot16=32$ states, $20N$ steps) for (a-d) $\omega=\left[1,1,1\right]$
(learning rate $\alpha=8\times10^{-7}$), (e-h) $\omega=\left[5,5,1\right]$
($\alpha=2.1\times10^{-5}$) and (i-l) $\omega=\left[1,1,b_{ii'}^{L}\right]$
($\alpha=2.1\times10^{-5}$). First and second columns show the weight
matrices before and after learning, respectively. Third column shows
energy at the end of convergence for a set without learning (resets
1-1000, blue), during learning (1001-2000, red), and after learning
(2001-3000, light blue). The right column shows the colored map resulting
from the learned state $\mathbf{S}$. Note that while the Hopfield
dynamics and learning are computed using modified weights for the
lower two rows, the energies in sub-plots g and k are computed from
the non-weighted constraints identical to sub-plot c for ease of comparison.\label{fig:3by4Plot}}
\end{figure*}

That said, given a situation when one is given just two colors when
four are required, we can still ask what is the optimal coloring scheme
to obtain a map as comprehensible as possible under those restrictions?

To answer this question we put additional weights $\omega_{i}$ on
the different set of constraints $\varphi_{i}$ in \eqref{eq:MapColor_cnf}:
\begin{align}
\Phi= & \omega_{1}\varphi_{1}\land\omega_{2}\varphi_{2}\land\omega_{3}\varphi_{3},\label{eq:MapColor_cnf-1}
\end{align}
and tested 3 different scenarios: 
\begin{enumerate}
\item $\omega=\left[1,1,1\right]$ - all constraints have the same weight. 
\item $\omega=\left[5,5,1\right]$ - the color constraints $\varphi_{1},\varphi_{2}$
have a higher weight compared to the set of border constraints $\varphi_{3}$.
Meaning that each region first and foremost must be a proper color. 
\item $\omega=\left[1,1,b_{ii'}^{L}\right]$ - the border constraints $\varphi_{3}$
are weighted by a normalized border adjacency matrix $B^{L}$ with
the elements $b_{ii'}^{L}$ denoting the border length between countries
$i$ and $i'$. 
\end{enumerate}
The initial weights $\mathbf{W}_{0}$ and weight matrices $\mathbf{W}$
after the SO simulation for each of these weighed sets are shown in
the first two columns of Fig.~\ref{fig:3by4Plot}. We can see from
Fig.~\ref{fig:3by4Plot} how the initial weights $\mathbf{W}_{0}$
change corresponding to the different additional weights $\omega$
on each of the set of constraints. The second column shows the weight
matrices $\mathbf{W}$ after learning (after the learning period shown
in red in the third column of Fig.~\ref{fig:3by4Plot}). From Fig.~\ref{fig:3by4Plot}
we can see several things. First, for the same amount of time, each
of the scenarios end in a different result. The lowest energy (corresponding
to the lowest number of broken clauses) is achieved for the $\omega=\left[1,1,1\right]$
scenario (Fig.~\ref{fig:3by4Plot}, c and d), which corresponds to
the regular unweighted formula $\Phi$. For comparison we checked
the CNF instance of the problem with the RC2 solver in the PySAT package
\cite{ignatiev_pysat_2018} and it produced the same result. This
result, however, shows that three countries do not have a proper color
(indicated by yellow color in Fig.~\ref{fig:3by4Plot}d). In comparison,
the additional weights on the constraints $\omega=\left[5,5,1\right]$
(Fig.~\ref{fig:3by4Plot}, second row) and $\omega=\left[1,1,b_{ii'}^{L}\right]$
(Fig.~\ref{fig:3by4Plot}, third row) produced proper colors for
all countries, which however results in a higher energy (higher number
of border constraints broken). Although, the two weights on the constraints
$\omega=\left[5,5,1\right]$ and $\omega=\left[1,1,b_{ii'}^{L}\right]$
serve the same function of making sure that all countries have proper
color, everything else taken equal, they produce different results.
An interesting result is obtained for $\omega=\left[1,1,b_{ii'}^{L}\right]$
(Fig.~\ref{fig:3by4Plot}, k and l). We can see that Costa Rica has
the same color as Panama, despite the fact that it does not cost anything
to make it a different color. In fact, making it a different color
would decrease the energy. But what happens here is that learning
is too fast, so it changes the energy landscape to the extent that
this constraint is broken ``forever''. The Costa Rica state is no
longer a local minimum for the original weight matrix. This is easy
to fix (see Appendix~\ref{sec:borderWeightFixed}) but provides valuable
insight into the interplay between learning and constraints as discussed
in the next section.

\section{Discussion\label{sec:Discussion}}

This work demonstrates the practical application of the Self-Optimization
(SO) algorithm to concrete combinatorial problems in SAT form. Through
two different examples, the Liars problem and the map coloring problem
of a checkerboard, we showcase the model's capability to find optimal
configurations that represent the solutions to these problems. Before
we discuss the example of coloring the map of South America, we want
to point out here the current limitations of the model.

In all the example problems used, there is a maximum $k=2$ literals
per clause. This type of problems is known as Max-2-SAT. The examples
of adding additional weights on the constraints in Sec.~\ref{subsec:South America}
can be viewed as weighted-Max-2-SAT problems. The current implementation
of the SO model cannot handle SAT instances with $k>2$ literals per
clause\footnote{For example, in the Liars problem, $k=3$ literals would be in the
case if a person made a statement about two people (e.g. "Cal says
Alice and Bob are both liars"). In the map coloring problem, that
would be the case if there would be three colors available.}. That said it is known that any k-SAT can be reduced to 3-SAT problem,
and any 3-SAT problem can be reduced to Max-2-SAT \cite{papadimitriou_computational_1994}
(in both cases at the expense of a linear number of new variables).
This means that, in principle, any k-SAT problem can be reduced to
Max-2-SAT and then the SO model can be used to solve it.

Another limitation of the model is its poor handling of problems without
an underlying structure. As we have seen in Sec.~\ref{subsec:Solvable-problems},
although the model does find the solution for the Liars problem, it
takes a considerable amount of time even though it was a small problem.
This becomes even worse for problems of bigger size. This is due to
a known limitation of selective associations \cite{watson_transformations_2011-1}.
For such problems, the stochastic nature of repeated resets also can
be problematic. The Liars Problem discussed in Sec~\ref{subsec:Solvable-problems},
for instance, requires fine tuning of the learning rate depending
on the seed of the random number generator to converge to the zero
energy configuration. Thus one should consider the structure of their
problem before using the SO model to solve it.

Despite these limitations, there are many benefits for using the SO
model. First, in comparison to existing SAT solvers, although the
SO model is not comparable speed-wise, different than SAT solvers
the SO model a) finds a solution in a biologically realistic way,
and b) it has potential for parallelization even though our implementation
is sequential. We note also that in the current work at each reset
the entire system was reset instantaneously. However, recently we
showed \cite{froese_autopoiesis_2023} that resetting just part of
the system allows the system to reach the lowest converged energy
in far fewer resets. This adds both more biological plausibility and
faster processing.

Moreover, even if for some combinatorial optimization problems the
SO model might not produce solutions that will satisfy all the constraints,
our experiments indicate that it can still yield high quality solutions.
These solutions prove beneficial in various contexts in optimization
solvers. For instance, such a high quality, but imperfect, solution
to a combinatorial optimization problem provides valuable insight
into the local consistencies of variable assignments. These assignments
could potentially be used to guide the search in a SAT solver, serving
as variable selection or phase selection heuristics\cite{biere_handbook_2009}.
Most modern solvers use tree search in order to guarantee algorithmic
completeness, though some also incorporate local search at various
stages of the process. In scenarios like those described in \cite{cai_better_2022},
the tree search can be halted at any point, allowing partial assignment
at that moment to be heuristically completed to a full assignment,
even if it means violating some constraints. Subsequently, local search
can be performed on the assignment, with attempts to mutate it into
a correct solution. This approach allows for alternative methods of
completing partial solutions into full solutions, as well as alternative
ways of conducting local search. We posit that SO can be successfully
utilized in these situations, and we plan to explore these possibilities
in future research.

Finally, we further expanded the approach by analyzing a more complex
problem: coloring the map of South America with just two colors. This
problem has no solution. In principle, four colors are needed to avoid
color clashes at borders. The goal of that example was to provide
insight into what happens to the end result when individual constraints
of the problem are broken.

Previous investigations into the properties of Hebbian learning in
the context of neural networks focused on abstract problems. This
provides a valuable basis for understanding the process and allows
for a thorough analysis of its behavior. We can add to that using
concrete problems by analyzing what it means to break constraints.
Specifically, the last example of Sec.~\ref{subsec:South America},
shows how Hebbian learning has dramatically modified the weight matrix
that encodes the problem constraints such that some of these constraints
are no longer contained within. In the example, in the unmodified
weight matrix the color of Costa Rica can be trivially changed to
satisfy the constraint on having opposite colors across borders, however
a state with that color flipped is no longer an energetic minimum
of the learned weight matrix. An observation of this form cannot be
derived from changes to the weight matrix in an abstract problem.
At this point we do not know how to exploit that observation to improve
the model or the approach, but this should definitely provide an avenue
for further research. For instance it is interesting to investigate
how breaking some constraints helps in solving the rest of the problem.
Another fruitful direction of inquiry might be to check whether broken
constraints can later be mended. A concrete problem can be utilized
in these cases to understand the effect of changes to the model or
the solution strategy.

\section{Data availability statement }

The model from the main text as well as the code used for the simulation
are available at \cite{weber_so_2023}. The shapes comprising the
map of South America in Sec.~\ref{subsec:South America} were obtained
from the CShapes 2.0 Dataset \cite{schvitz_mapping_2022}.

\section{Acknowledgment }

We are grateful for the help and support provided by the Scientific
Computing and Data Analysis section of Research Support Division at
OIST.

\appendix

\section{Lowered learning rates for border-length weights}

\label{sec:borderWeightFixed} 
\begin{figure}
\includegraphics{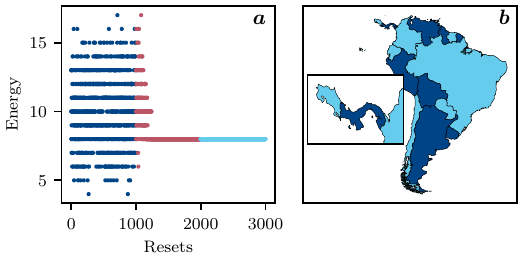}

\caption{Self-optimization simulation results for coloring the map of South
America ($N=2\cdot16$ states, $20N$ steps) $\omega=\left[1,1,b_{ii'}^{L}\right]$
($\alpha=6\times10^{-6}$). Sub-plot a shows energy at the end of
convergence for a set without learning (resets 1-1000, blue), during
learning (1001-2000, red), and after learning (2001-3000, light blue).
Subplot b shows the colored map resulting from the learned state $\mathbf{S}$.
For this learning rate, the weights on the constraints $\omega=\left[1,1,b_{ii'}^{L}\right]$
result in the system converging to the same energy state as for $\omega=\left[5,5,1\right]$
(compare to Fig.~\ref{fig:3by4Plot}g-h). \label{fig:1by2Extra}}
\vspace*{-3mm}
 
\end{figure}

The broken border color constraint
resulting from constraint weighting by border length as depicted in
Fig.~\ref{fig:3by4Plot} sub-plot l can be fixed by employing a less
aggressive learning rate as shown in Fig.~\ref{fig:1by2Extra}. 

\bibliographystyle{IEEEtran}
\bibliography{SO_for_SAT}

\end{document}